\documentclass[12pt,tightenlines]{article}
\usepackage{rotating}
\usepackage{mathptmx}
\usepackage{geometry}
\usepackage{titling}
\usepackage{authblk}
\usepackage{amssymb}
\usepackage{amsmath}
\usepackage{amsthm}
\allowdisplaybreaks[4]
\usepackage{slashed}
\usepackage{graphicx,color,epsfig}
\usepackage{multirow}
\usepackage{cite}
\usepackage[colorlinks=true,linkcolor=red,citecolor=blue]{hyperref}
\newcommand{\beq}{\begin{eqnarray}}
\newcommand{\eeq}{\end{eqnarray}}

\begin{document}
\title{Quasi-Two-Body $B\to P (K^+K^-, \pi^+\pi^-)$ Decays with $f_0(980)$ Resonance}
\author{Ya-Xin Wang}
\author{Jun-Peng Wang}
\author{Ying Li$\footnote{liying@ytu.edu.cn}$}
\author{Zhi-Tian Zou$\footnote{zouzt@ytu.edu.cn}$}
\affil{\it Department of Physics, Yantai University, Yantai 264005, China}
\maketitle
\vspace{0.2cm}

\begin{abstract}
In this study, we investigate the $CP$-averaged branching fractions and direct $CP$ asymmetries in the quasi-two-body decays $B\to P (f_0(980) \to )(K^+K^-, \pi^+\pi^-)$ within the framework of perturbative QCD approach, where $P$ denotes light pseudoscalar mesons ($\pi$, $K$). The timelike form factor $F_{S}(\omega)$, characterizing final-state interactions between collinear particles in the resonant region, is modeled through the revised relativistic Breit-Wigner formalism for the $S$-wave $f_0(980)$ resonance. Within the Gegenbauer moments of the $S$-wave two-meson distribution amplitudes constrained by current data, we predict the branching fractions of $B\to P (f_0(980) \to )(K^+K^-, \pi^+\pi^-)$ to lie in the range of $10^{-8}$ to $10^{-6}$.  Our calculations reveal that our theoretical results for some decay modes exhibit satisfactory agreement with existing measurements from the BaBar and Belle collaborations. Furthermore, we extract the corresponding branching fractions of  $B \to P f_0(980)$ decays from the quasi-two-body counterparts $B\to P (f_0(980) \to )\pi^+\pi^-$ rather than  $B\to P (f_0(980) \to )K^+K^-$. All predictions are awaiting experimental verification in current and upcoming collider experiments.
\end{abstract}

\newpage
\section{Introduction}
Hadronic decays of the $B$ meson provide crucial insights into three key areas: $CP$ violation and the determination of the Cabibbo-Kobayashi-Maskawa (CKM) matrix angles \cite{Kobayashi:1973fv}, the structure of QCD in the presence of heavy quarks and energetic light particles, and the potential exploration of new physics beyond the standard Model (SM) in the quark sector. Over the past few decades, experimental data on non-leptonic $B$ decays has been steadily accumulated from various experiments, including CLEO, the $B$-factories BaBar and Belle \cite{BaBar:2014omp}, the Tevatron \cite{Kuhr:2013hd}, and LHCb \cite{LHCb:2012myk}. Furthermore, the Belle II experiment is poised to conduct high-precision analyses of non-leptonic $B$ decays \cite{Belle-II:2018jsg}. 

Due to significant combinatorial backgrounds, research on charmless $B$ decays has largely focused on two-body decay processes. However, in recent years, there has been growing interest in three-body $B$ decays, both experimentally and theoretically (for reviews, see \cite{Virto:2016fbw, Liu:2021sdw}). In contrast to two-body decays, where the kinematics are completely determined, the amplitude of a three-body decay depends on two invariant masses (e.g., $s_{12}$ and $s_{13}$), defined as $s_{ij} \equiv (p_i + p_j)^2$. The allowed kinematic configurations define a triangular region in the $s_{12}$-$s_{13}$ plane. This region is commonly represented by a Dalitz plot \cite{Dalitz:1953cp}, a powerful tool widely used in both experimental and theoretical analyses of three-body decays. The Dalitz plot allows us to distinguish between resonant and non-resonant contributions, offering valuable insights into the dynamics of the decay process. Typically, the Dalitz plot is divided into distinct regions based on the kinematic characteristics of the decay. The central region corresponds to non-resonant contributions, where the three final-state particles are emitted with roughly equal energy ($E \sim m_B / 3$). The three corners of the plot correspond to kinematics in which one final particle is nearly at rest, while the other two move back-to-back with energy $E \sim m_B / 2$. The edges of the Dalitz plot represent cases where two final-state particles are collinear, with the third particle recoiling. In these edges, the two collinear particles may originate from an intermediate resonance. Studying the edges of the Dalitz plot is crucial for probing the properties of these resonances.

Given the significant physical importance of studying three-body decays of $B$ mesons, several theoretical methods have been developed to investigate this phenomenon. These include QCD factorization \cite{El-Bennich:2009gqk, Krankl:2015fha, Klein:2017xti, Cheng:2016shb, Cheng:2014uga, Li:2014oca, Lesniak:2024zrc}, the perturbative QCD (PQCD) approach \cite{Wang:2016rlo, Li:2016tpn, Rui:2017bgg, Zou:2020atb, Zou:2020fax, Zou:2020ool}, and symmetry-based methods \cite{Zhang:2013oqa, El-Bennich:2006rcn, Hu:2022eql, Abreu:2023hts}. These approaches offer complementary tools for understanding the complex dynamics of three-body decays, providing valuable insights into both the structure of resonances and potential contributions from final-state interactions.

These decays not only offer a broader understanding of $B$ meson decay mechanisms but also provide valuable insights into the nature of the resonances involved, many of which are still not fully understood. For example, the complex quark structure of light hadrons has remained a topic of active debate for years \cite{ParticleDataGroup:2024cfk}, despite the notable success of the constituent quark model. Specifically, $f_0(980)$ is a scalar particle with a mass around $980~{\rm MeV}$, and it has been the subject of extensive research due to its peculiar properties and its role in hadronic physics \cite{Achasov:2020fee}. Its nature remains an open question, with various competing models proposed to explain its characteristics, such as being a conventional quark-antiquark state \cite{Close:2002zu}, a meson-meson molecule \cite{Weinstein:1990gu}, or a tetraquark state \cite{Achasov:2020aun}. In three-body decays like $B \to P (f_0(980) \to) P (K^+K^-, \pi^+\pi^-)$, the $f_0(980)$ acts as an intermediate state, and its contributions to the decay amplitude can reveal crucial information about its properties. The decays of $B$ mesons into final states involving this resonance provide a direct probe of the internal structure and dynamics of the $f_0(980)$.

In recent years, measurements of various decays, including $B \to \pi\pi\pi$, $B \to K\pi\pi$, $B \to \pi KK$, and $B \to KKK$, by the BaBar \cite{BaBar:2005qms, BaBar:2007itz, BaBar:2008lpx, BaBar:2009jov, BaBar:2009vfr, BaBar:2011ktx, BaBar:2011vfx, BaBar:2012iuj}, Belle \cite{Belle:2004drb, Belle:2005rpz, Belle:2006ljg, Belle:2008til, Belle:2010wis, Belle:2017cxf}, and LHCb \cite{LHCb:2013dkk, LHCb:2013lcl, LHCb:2013ptu, LHCb:2014mir, LHCb:2019jta, LHCb:2019sus, LHCb:2019vww, LHCb:2019xmb, LHCb:2022fpg} collaborations have sparked significant theoretical interest in understanding three-body hadronic $B$ meson decays and exploring the nature of light hadrons. Experimentally, these decays are known to be dominated by low-energy resonances in the $\pi\pi$ and $KK$ channels. In the invariant mass distribution of $\pi^+\pi^-$ (or $\pi^0\pi^0$), a cusp-like structure followed by a disappearance around 1 GeV is observed, which may be associated with the scalar meson $f_0(980)$, while an enhancement near the $K^+K^-$ (or $\overline{K}K$) threshold has also been found. Motivated by these observations, we shall investigate the edges of the Dalitz plot for the processes $B \to \pi\pi\pi$, $B \to K\pi\pi$, $B \to \pi KK$, and $B \to KKK$ decays within the PQCD approach, aiming to probe potential insights into the nature of the $f_0(980)$ and to test the applicability of the PQCD approach in calculating three-body decays of $B$ mesons.

The present paper is organized as follows. In Sec.~\ref{sec:function}, we give a brief introduction for the theoretical framework. The numerical results and some discussions  will be given in Sec.~\ref{sec:result}. At last, we summarize this work in Sec.~\ref{sec:summary}.
\section{The Decay Formalism and Wave Functions}\label{sec:function}
The description of three-body $B$ meson decays remains largely model-dependent, with approaches such as the isobar model \cite{Herndon:1973yn} and the K-matrix formalism  \cite{Chung:1995dx} playing a central role. These models effectively capture the resonant contributions, while non-resonant contributions are often described using empirical distributions. In experimental and theoretical analyses, the isobar model, commonly applied to the Dalitz plot, represents the total decay amplitude as a coherent sum of amplitudes from $N$ individual decay channels:
\begin{eqnarray}
\mathcal{A}=\sum_{j=1}^{N}c_j\mathcal{A}_j,
\end{eqnarray}
where $c_j$ are complex coefficients that determine the relative magnitude and phase of each decay channel. Unlike two-body $B$ meson decays, the interference between different channels could introduces a new source of $CP$ asymmetry in the SM.

For the $B \to P(K^+K^-,\pi^+\pi^-)$ decays involving the $f_0(980)$ resonance, the resonant decay amplitude $\mathcal{A}_j$ can be calculated perturbatively within the PQCD approach. The resonant contributions are associated with the edges of the Dalitz plot, where two particles move collinearly with large energy, while the bachelor meson also recoils with high energy. The low-energy interactions between the two particles flying collinearly give rise to resonance formation. Therefore, these decays are similar to two-body decays, but with two meson pairs instead of a single meson.  Conceptually, the factorization approach for two-body decays can also be extended to handle quasi-two-body $B$ meson decays involving an intermediate resonance. However, this extension requires more complex hadronic inputs, such as a two-meson distribution amplitude. In the PQCD framework, the decay amplitude of the quasi-two-body decay can be decomposed as a convolution:
\begin{eqnarray}
\mathcal{A}\sim\Phi_B\otimes \Phi_{M_1}\otimes\Phi_{M_2M_3}\otimes H, \label{fq}
\end{eqnarray}
where $H$ is the hard kernel, which is perturbatively calculable. $\Phi_B$ is the wave function of the $B$ meson, and $\Phi_{M_1}$ is the wave function of the bachelor meson. The new introduced $\Phi_{M_2M_3}$ is the two-meson wave function, which represents the new, complicated hadronic input.  It is important to note that the two-meson wave function has not yet been studied from first principles, so it can only be determined phenomenologically at present.

To perturbatively calculate the decay amplitude $\mathcal{A}_j$ for $B \to P(K^+K^-,\pi^+\pi^-)$ decays involving the $f_0(980)$ resonance, the first step is to present the effective Hamiltonian for the weak interaction, which governs the hard kernel $H$. The effective Hamiltonian can be written as \cite{Buchalla:1995vs}:
\begin{eqnarray}
\mathcal{H}_{eff}=\frac{G_F}{\sqrt{2}}\bigg\{V_{ub}^*V_{ud(s)}(C_1O_1+C_2O_2) 
-V_{tb}^*V_{td(s)}\sum_{i=3}^{10}C_iO_i\bigg\},
\end{eqnarray}
where $V_{UD}$ are the relevant CKM matrix elements. The operators $O_{1-10}(\mu)$ represent local four-quark operators that describe the quasi-two-body $B \to P(K^+K^-,\pi^+\pi^-)$ decays, and they are grouped into three distinct categories:
\begin{itemize}
    \item Current-Current Operators
    \begin{eqnarray}
O_1&=&(\bar{b}_{\alpha}u_{\beta})_{V-A}(\bar{u}_{\beta}q_{\alpha})_{V-A}, \\
O_2&=&(\bar{b}_{\alpha}u_{\alpha})_{V-A}(\bar{u}_{\beta}q_{\beta})_{V-A}, 
\end{eqnarray}
with $q=d,s$ quark,
\item QCD Penguin Operators
\begin{eqnarray}
O_3&=&(\bar{b}_{\alpha}q_{\alpha})_{V-A}\sum_{q^{\prime}}(\bar{q}^{\prime}_{\beta}q^{\prime}_{\beta})_{V-A}, \\
O_4&=&(\bar{b}_{\alpha}q_{\beta})_{V-A}\sum_{q^{\prime}}(\bar{q}^{\prime}_{\beta}q^{\prime}_{\alpha})_{V-A}, \\
O_5&=&(\bar{b}_{\alpha}q_{\alpha})_{V-A}\sum_{q^{\prime}}(\bar{q}^{\prime}_{\beta}q^{\prime}_{\beta})_{V+A}, \\
O_6&=&(\bar{b}_{\alpha}q_{\beta})_{V-A}\sum_{q^{\prime}}(\bar{q}^{\prime}_{\beta}q^{\prime}_{\alpha})_{V+A}, 
\end{eqnarray}
\item Electroweak Penguin Operators
\begin{eqnarray}
O_7&=&\frac{3}{2}(\bar{b}_{\alpha}q_{\alpha})_{V-A}\sum_{q^{\prime}}e_{q^{\prime}}(\bar{q}^{\prime}_{\beta}q^{\prime}_{\beta})_{V+A}, \\
O_8&=&\frac{3}{2}(\bar{b}_{\alpha}q_{\beta})_{V-A}\sum_{q^{\prime}}e_{q^{\prime}}(\bar{q}^{\prime}_{\beta}q^{\prime}_{\alpha})_{V+A}, \\
O_9&=&\frac{3}{2}(\bar{b}_{\alpha}q_{\alpha})_{V-A}\sum_{q^{\prime}}e_{q^{\prime}}(\bar{q}^{\prime}_{\beta}q^{\prime}_{\beta})_{V-A}, \\
O_{10}&=&\frac{3}{2}(\bar{b}_{\alpha}q_{\beta})_{V-A}\sum_{q^{\prime}}e_{q^{\prime}}(\bar{q}^{\prime}_{\beta}q^{\prime}_{\alpha})_{V-A}. 
\end{eqnarray}
\end{itemize}
The subscripts $\alpha$ and $\beta$ represent the color indices. The active quarks $q^{\prime}$ at the scale $m_b$, with charge $e_{q^{\prime}}$, include $q^{\prime} = u, d, s$ quarks in this work. The left-handed current $(\bar{b}_{\alpha}q_{\alpha})_{V-A}$ is given by the explicit expression $\bar{b}_{\alpha}\gamma_{\mu}(1-\gamma_5)q_{\alpha}$, while the right-handed current $(\bar{q}^{\prime}_{\beta}q^{\prime}_{\beta})_{V+A}$ is $\bar{q}^{\prime}_{\beta}\gamma_{\mu}(1+\gamma_5)q^{\prime}_{\beta}$. For convenience, the combined Wilson coefficients $a_i$ are defined as:
\begin{eqnarray}
&&a_1=C_2+C_1/3,\nonumber\\
&&a_2=C_1+C_2/3,\nonumber\\
&&a_i=C_i+C_{i+1}/3,\;\, i=3,5,7,9; \nonumber\\
&&a_j=C_j+C_{j-1}/3.\;\, j=4,6,8,10.
\end{eqnarray}

The wave functions in the factorization formula of eq.~\eqref{fq} serve as crucial inputs in the PQCD approach, directly influencing the precision of the theoretical predictions. Therefore, employing the appropriate wave functions is of importance in our calculations. The wave function of the $B$ mesons, $\Phi_B$, and the wave function $\Phi_{M_1}$ for the light bachelor mesons ($K$ and $\pi$) have been well-determined and are widely used in analyses of two-body $B \to PP, PV, VV$ decays \cite{Lu:2000em, Yu:2005rh, Ali:2007ff, Zou:2015iwa}. Since these are already well-established, we will not revisit them in this work. However, the two-meson wave functions $\Phi_{M_2M_3}$, which represent new and more complicated hadronic inputs, must be discussed in detail. To analyze the three-body $B \to P(K^+K^-,\pi^+\pi^-)$ decays with the $f_0(980)$ scalar resonance as the intermediate state, appropriate $S$-wave two-meson wave functions are required for both the $KK$ and $\pi\pi$ pairs. It's important to note that the theoretical description of the two-meson wave function is still in the modeling phase. This means that accurate wave functions are not yet available from a QCD-inspired approach, and their forms are instead determined phenomenologically, with the parameters constrained by matching available experimental measurements. For the $S$-wave two-meson wave function, following the phenomenological analysis in \cite{Li:2019hnt, Rui:2019yxx, Wang:2015uea, Xing:2019xti}, it can be expanded as:
\begin{eqnarray}
\Phi_S=\frac{1}{2\sqrt{N_c}}\bigg[P\mkern-10.5mu/\phi_S(z,\zeta,\omega)+\omega\phi_S^s(z,\zeta,\omega)
+\omega(n\mkern-10.5mu/ v\mkern-10.5mu/-1)\phi_S^t(z,\zeta,\omega)\bigg],
\end{eqnarray}
where $P$ and $\omega$ denote the momentum and the invariant mass of the two-meson pair, respectively, with $P^2 = \omega^2$. The vectors $n = (1, 0, \mathbf{0}_T)$ and $v = (0, 1, \mathbf{0}_T)$ are dimensionless light-like vectors. The functions $\phi_S$ and $\phi_S^{s,t}$ represent the twist-2 and twist-3 light-cone distribution amplitudes (LCDAs), respectively. The parameter $z$ is the momentum fraction of the spectator quark, and $\xi$ is the momentum fraction of one meson in the two-meson pair.

Following the expansion of the LCDA for mesons, the LCDAs for the two-meson pair can also be expressed in terms of Gegenbauer polynomials with their corresponding Gegenbauer moments:
\begin{eqnarray}
&& \phi_S(z,\xi,\omega)=\frac{6}{2\sqrt{2N_C}}F_S(\omega)z(1-z)\left[B_1C_1^{3/2}(1-2z)+B_3C_3^{3/2}(1-2z)\right],\\
&& \phi_S^s(z,\xi,\omega)=\frac{1}{2\sqrt{2N_c}}F_S(\omega),\\
&& \phi_S^t(z,\xi,\omega)=\frac{1}{2\sqrt{2N_C}}F_S(\omega)(1-2z),
\end{eqnarray}
where $C_{1,3}^{3/2}(1 - 2z)$ are the Gegenbauer polynomials, and $B_{1,3}$ represent the corresponding Gegenbauer moments. It should be noted that the twist-3 LCDAs provided here are set to be the asymptotic forms, which are sufficiently accurate for this analysis. As aforementioned, the study of the two-meson wave function is still in the phenomenological stage, and  the values of the Gegenbauer moments $B_{1,3}$ are determined through a combination of theoretical modeling and current experimental data. In this work, we adopt $B_1 = -0.8$ and $B_3 = 0.2$ for $KK$ pair\cite{Zou:2020atb,Li:2020zng}, and $B_1 = -0.8$ and $B_3 = 0.72$ for  $\pi\pi$ pair \cite{Jia:2021uhi}. Notably, the values of the Gegenbauer moments for the $KK$ pair differ significantly from those for the $\pi\pi$ pair. This difference arises because the $K$ meson is more massive than the $\pi$ meson. When the combined mass of the two $K$ mesons approaches the mass of the intermediate resonance, the behavior of the wave function is substantially affected, particularly for the $f_0(980)$ resonance. In addition, in Ref. \cite{Zou:2020atb}, we had investigated the quasi-two-body $B \to KKK$ decays with the intermediate $f_0(980)$ resonance, where we approximated by neglecting both the mass of the final $K$ meson and the $C_3^{3/2}(1-2z)$ term in the twist-2 LCDA $\phi_S(z, \xi, \omega)$. In the current work, we update this analysis by including the mass of the $K$ meson and incorporating a more accurate treatment of the LCDAs for the $KK$ pair, specifically including the $C_3^{3/2}(1-2z)$ term.

In contrast to the LCDA of meson, the time-like form factor $F_S(\omega)$ is introduced in our description of LCDA of meson-pair. It is important to note that the interpretation of this form factor is still in the modeling stage. The relativistic Breit-Wigner (RBW) model \cite{ParticleDataGroup:2018ovx} is widely used in experimental analyses and theoretical descriptions and has been successful for resonances with narrow widths. However, the RBW model fails to adequately describe $F_S(\omega)$ for the $f_0(980)$ resonance, as it exhibits an anomalous structure near $980~{\rm MeV}$, corresponding to an enhancement from the $KK$ system observed in $\pi\pi$ scattering \cite{Alston-Garnjost:1971lsd, Flatte:1972rz}. This structure can be interpreted as a resonance formed by a combination of two channels. Consequently, the Flatt$\acute{e}$ form has been proposed to better describe the time-like form factor of the $f_0(980)$ resonance \cite{Flatte:1976xu, Bugg:2008ig, LHCb:2014ooi}. For the Flatt$\acute{e}$ form, we adopt the updated expression from Ref. \cite{LHCb:2014ooi}:
\begin{eqnarray}
F_S(\omega)=\frac{m^2_{f_0}}{m^2_{f_0}-\omega^2-im_{f_0}(g_{\pi\pi}\rho_{\pi\pi}+g_{KK}\rho_{KK}F_{KK})}.
\end{eqnarray}
where $g_{\pi\pi}$ and $g_{KK}$ are the coupling constants of the $f_0(980)$ to the $\pi\pi$ and $KK$ channels, respectively. The phase space factors $\rho_{\pi\pi}$ and $\rho_{KK}$ are typically modeled as:
\begin{eqnarray}
\rho_{\pi\pi}=\sqrt{1-\frac{4m_{\pi}^2}{\omega^2}},\;\;\rho_{KK}=\sqrt{1-\frac{4m_{K}^2}{\omega^2}}.
\end{eqnarray}
The factor $F_{KK} = e^{-\alpha q^2}$, introduced in Ref. \cite{Bugg:2008ig}, is used to suppress the $KK$ contribution, with the parameter $\alpha \approx 2.0~{\rm GeV}^{-2}$ and $q$ being the momentum of each kaon in the $KK$ rest frame.

\begin{figure*}[!htb]
\begin{center}
\includegraphics[scale=1.25]{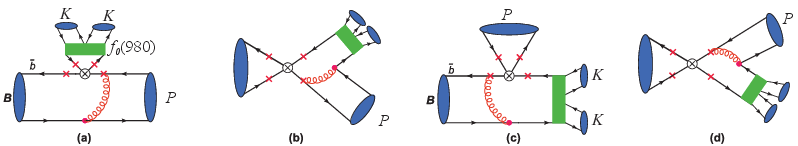}
\caption{Typical Feynman diagrams for the quasi-two-body decays $B\to Pf_0(980)\to PK^+K^-$ in PQCD,
in which the symbol $\otimes$ stands for the weak vertex, $\times$ denotes
possible attachments of hard gluons, and the green rectangle represents intermediate
states $f_0(980)$.}\label{feynman}
\end{center}
\end{figure*}
In Figure.~\ref{feynman}, we present the Feynman diagrams contributing to the quasi-two-body decay at the leading order in PQCD approach, by taking $B\to P (f_0(980) \to )K^+K^-$ as an example. By applying the factorization formula and using the wave functions of both the initial and final states, we can analytically calculate the decay amplitudes, and they are expressed as follows: 
{\small
\begin{eqnarray}
\mathcal{A}[B^0\to K^0(f_0(q\bar{q})\to)PP]&=&\frac{G_F}{2}\Bigg\{V_{ub}^*V_{us}\Bigg[C_2M_{K}^{LL}\Bigg]-V_{tb}^*V_{ts}\Bigg[
 \left(a_4-\frac{a_{10}}{2}\right)F_{PP}^{LL}
+\left(a_6-\frac{a_8}{2}\right)F_{PP}^{SP}\nonumber\\
&&
+\left(2C_4+\frac{C_{10}}{2}\right)M_{K}^{LL}
+\left(2C_6+\frac{C_{8}}{2}\right)M_{K}^{SP}
+\left(C_3-\frac{C_{9}}{2}\right)M_{PP}\nonumber\\
&& 
+\left(C_5-\frac{C_{7}}{2}\right)M_{PP}^{LR}
+\left(a_4-\frac{a_{10}}{2}\right)A_{PP}^{LL}
+\left(a_4-\frac{a_{8}}{2}\right)A_{PP}^{SP}\nonumber\\
&& 
+\left(C_3-\frac{C_{9}}{2}\right)W_{PP}^{LL}
+\left(C_5-\frac{C_{7}}{2}\right)W_{PP}^{LR}\Bigg]\Bigg\},\\
\mathcal{A}[B^0\to K^0(f_0(s\bar{s})\to)PP]&=&-\frac{G_F}{\sqrt{2}}V_{tb}^*V_{ts}\Bigg[
\left(a_6-\frac{a_8}{2}\right)F_{K}^{SP}
+\left(C_3+C_4-\frac{C_9}{2}-\frac{C_{10}}{2}\right)M_{K}^{LL}\nonumber\\
&&
+\left(C_5-\frac{C_{7}}{2}\right)M_{K}^{LR}
+\left(C_6-\frac{C_8}{2}\right)M_{K}^{SP}
+\left(a_4-\frac{a_{10}}{2}\right)A_{K}^{LL}\nonumber\\
&&
+\left(a_6-\frac{a_8}{2}\right)A_{K}^{SP}
+\left(C_3-\frac{C_9}{2}\right)W_{K}^{LL}
+\left(C_5-\frac{C_7}{2}\right)W_{K}^{LR}\Bigg],\\
\mathcal{A}[B^0\to \pi^0(f_0(q\bar{q})\to)PP]&=&\frac{G_F}{2\sqrt{2}}\Bigg\{V_{ub}^*V_{ud}\Bigg[
a_2\left(F_{PP}^{LL}
+A_{\pi}^{LL}+A_{PP}^{LL}\right)
+C_2\left(M_{PP}^{LL}-M_{\pi}^{LL}+W_{\pi}^{LL}+W_{PP}^{LL}\right)\Bigg]\nonumber\\
&& -V_{tb}^*V_{td}\Bigg[
 \left(-a_4-\frac{3a_7}{2}+\frac{4C_9}{3}+a_{10}\right)F_{PP}^{LL}
+\left(\frac{a_8}{2}-a_6\right)F_{PP}^{SP} \nonumber\\
&& 
-\left(a_6-\frac{a_8}{2}\right)F_{\pi}^{SP}
-\left(C_3+2C_4-\frac{C_9}{2}+\frac{C_{10}}{2} \right)M_{\pi}^{LL} 
-\left(C_5-\frac{C_7}{2}\right)M_{\pi}^{LR}\nonumber\\
&& 
-\left(2C_6+\frac{C_8}{2}\right)M_{\pi}^{SP}
+\left(\frac{3a_{10}}{2}-C_3\right)M_{PP}^{LL} 
+\left(\frac{C_{7}}{2}-C_5\right)M_{PP}^{LR}\nonumber\\
&& 
+\frac{3C_{8}}{2}M_{PP}^{SP}
+\left(-\frac{a_4}{2}-\frac{3a_7}{4} +\frac{5C_9}{6}+\frac{C_{10}}{2}\right)\left(A_{\pi}^{LL}+A_{PP}^{LL}\right)  \nonumber\\
&&+\left(\frac{a_{8}}{4}-\frac{a_6}{2}\right)\left(A_{\pi}^{SP}+A_{PP}^{SP}\right)
+\left(\frac{3a_{10}}{4}-\frac{C_3}{2}\right)\left(W_{\pi}^{LL}+W_{PP}^{LL}\right) \nonumber\\
&&
+\left(C_7-\frac{C_5}{2} \right) \left(W_{\pi}^{LR}+W_{PP}^{LR}\right) 
+ \frac{3C_8}{4}\left(W_{\pi}^{SP}+W_{PP}^{SP}\right)\Bigg]\Bigg\},\\
\mathcal{A}[B^0\to \pi^0(f_0(s\bar{s})\to)PP]&=&-\frac{G_F}{2}V_{tb}^*V_{td}\left[\left(C_4-\frac{C_{10}}{2}\right)M_{\pi}^{LL}+\left(C_6-\frac{C_{8}}{2}\right)M_{\pi}^{SP}\right],\\
\mathcal{A}[B^+\to \pi^+(f_0(q\bar{q})\to)PP]&=&\frac{G_F}{2}\Bigg\{V_{ub}^*V_{ud}\Bigg[
 a_1\left(F_{PP}^{LL}+A_{PP}^{LL}+A_{\pi}^{LL}\right)
+C_1\left(M_{PP}^{LL}+W_{PP}^{LL}+W_{\pi}^{LL}\right)\nonumber\\
&&
+C_2 M_{\pi}^{LL}\Bigg]
-V_{tb}^*V_{td}\Bigg[
 \left(a_4+a_{10}\right)F_{PP}^{LL}
+\left(a_6+a_8\right)F_{PP}^{SP}
+\left(a_6-\frac{a_8}{2}\right)F_{\pi}^{SP}\nonumber\\
&&
+\left(C_3+C_9\right)M_{PP}^{LL}
+\left(C_5+C_7\right)M_{PP}^{LR}
+\left(C_3+2C_4-\frac{C_9}{2}-\frac{C_{10}}{2}\right)M_{\pi}^{LL}\nonumber\\
&&
+\left(C_5-\frac{C_{7}}{2}\right)M_{\pi}^{LR}
+\left(2C_6+\frac{C_8}{2}\right)M_{\pi}^{SP}
+\left(a_4+a_{10}\right)\left(A_{PP}^{LL}+A_{\pi}^{LL}\right)\nonumber\\
&&
+\left(a_6+a_8\right)\left(A_{PP}^{SP}+A_{\pi}^{SP}\right)
+\left(C_3+C_9\right)\left(W_{PP}^{LL}+W_{\pi}^{LL}\right)\nonumber\\
&&
+\left(C_5+C_7\right)\left(W_{PP}^{LR}+W_{\pi}^{LR}\right)\Big]\Bigg\},\\
\mathcal{A}[B^+\to \pi^+(f_0(s\bar{s})\to)PP]&=&-\frac{G_F}{\sqrt{2}}V_{tb}^*V_{td}\Bigg[
\left(C_4-\frac{C_{10}}{2}\right)M_{\pi}^{LL}
+\left(C_6-\frac{C_{8}}{2}\right)M_{\pi}^{SP}\Bigg],\\
\mathcal{A}[B^+\to K^+(f_0(q\bar{q})\to)PP]&=&\frac{G_F}{2}\Bigg\{V_{ub}^*V_{us}\Bigg[
 a_1\left(F_{PP}^{LL}+\mathcal{A}_{PP}^{LL}\right)
+C_1\left(M_{PP}^{LL}+W_{PP}^{LL}\right)
+C_2M_{K}^{LL}\Bigg]\nonumber\\
&&-V_{tb}^*V_{ts}\Bigg[
  \left(a_4+a_{10}\right)\left(F_{PP}^{LL}+A_{PP}^{LL}\right)
+\left(a_6+a_8\right)\left(F_{PP}^{SP}+A_{PP}^{SP}\right) \nonumber\\
&& 
+(C_3+C_9)\left(M_{PP}^{LL}+W_{PP}^{LL}\right)
+(C_5+C_7)\left(M_{PP}^{LR}+W_{PP}^{LR}\right)\nonumber\\
&& 
+\left(2C_4+\frac{C_{10}}{2}\right)M_{K}^{LL}
+\left(2C_6+\frac{C_{8}}{2}\right)M_{K}^{SP}\Bigg]\Bigg\},\\
\mathcal{A}[B^+\to K^+(f_0(s\bar{s})\to)PP]&=&\frac{G_F}{\sqrt{2}}\Bigg\{V_{ub}^*V_{us}\Bigg[a_1A_{K}^{LL}+C_1W_{K}^{LL}\Bigg]
-V_{tb}^*V_{ts}\Bigg[\left(a_6-\frac{a_8}{2}\right)F_{K}^{SP}\nonumber\\
&&
+\left(C_3+C_4-\frac{C_9}{2}-\frac{C_{10}}{2}\right)M_{K}^{LL}
+\left(C_5-\frac{C_7}{2}\right)M_{K}^{LR}\nonumber\\
&&
+\left(C_6-\frac{C_8}{2}\right)M_{K}^{SP}
+\left(a_4+a_{10}\right)A_{K}^{LL}
+\left(a_6+a_8\right)A_{K}^{SP}\nonumber\\
&&\left.+\left(C_3+C_9\right)W_{K}^{LL}+\left(C_5+C_7\right)M_{K}^{LR}\right]\Bigg\}\\
\mathcal{A}[B_s\to \pi^0(f_0(q\bar{q})\to)PP]&=&\frac{G_F}{2\sqrt{2}}\Bigg\{V_{ub}^*V_{us}\Bigg[a_2(A_{\pi}^{LL}+A_{PP}^{LL})
+C_2(W_{\pi}^{LL}+W_{PP}^{LL})\Bigg]\nonumber\\
&&
-V_{tb}^*V_{ts}\Bigg[\left(\frac{3C_9}{2} +\frac{C_{10}}{2}\right) \left(A_{\pi}^{LL}+A_{PP}^{LL}\right)
+\left(\frac{3C_7}{2} +\frac{C_{8}}{2}\right) \left(A_{\pi}^{LR}+A_{PP}^{LR}\right)\nonumber\\
&&
+\frac{3C_{10}}{2}\left(W_{\pi}^{LL}+W_{PP}^{LL}\right)
+\frac{3C_{8}}{2}\left(W_{\pi}^{SP}+W_{PP}^{SP}\right)\Bigg]\Bigg\},\\
\mathcal{A}[B_s\to \pi^0(f_0(s\bar{s})\to)PP]&=&\frac{G_F}{2}\Bigg\{V_{ub}^*V_{us}\left[a_2F_{PP}^{LL}+C_2M_{PP}^{LL}\right]-V_{tb}^*V_{ts}\Bigg[\frac{3C_8}{2}M_{PP}^{SP}
+\frac{3C_{10}}{2}M_{PP}^{LL}
\nonumber\\
&&+\left(\frac{3C_9}{2}+\frac{C_{10}}{2}-\frac{3C_7}{2}
-\frac{C_{8}}{2}\right)F_{PP}^{LL}
\Bigg]\Bigg\},\\
\mathcal{A}[B_s\to \bar{K}^0(f_0(q\bar{q})\to)PP]&=&\frac{G_F}{2}\Bigg\{V_{ub}^*V_{ud}\left[C_2M_{K}^{LL}\right]
-V_{tb}^*V_{td}\Bigg[\left(a_6-\frac{a_8}{2}\right)F_{K}^{SP}\nonumber\\
&&
+\left(C_3+2C_4-\frac{C_9}{2}-\frac{C_{10}}{2}\right)M_{K}^{LL}
+\left(C_5-\frac{C_7}{2}\right)M_{K}^{LR}\nonumber\\
&&
+(2C_6+\frac{C_8}{2})M_{K}^{SP}
+\left(a_4-\frac{a_{10}}{2}\right)A_{K}^{LL}
+\left(a_6-\frac{a_{8}}{2}\right)F_{K}^{SP}\nonumber\\
&&+\left(C_3-\frac{C_9}{2}\right)W_{K}^{LL}+\left(C_5-\frac{C_7}{2}\right)W_{K}^{LR}\Bigg]\Bigg\},\\
\mathcal{A}[B_s\to \bar{K}^0(f_0(s\bar{s})\to)PP]&=&-\frac{G_F}{\sqrt{2}}V_{tb}^*V_{td}\Bigg[
\left(a_4-\frac{a_{10}}{2}\right)F_{PP}^{LL}
+\left(a_6-\frac{a_8}{2}\right)F_{PP}^{SP}
+\left(C_3-\frac{C_9}{2}\right)M_{PP}^{LL}\nonumber\\
&&
+\left(C_5-\frac{C_7}{2}\right)M_{PP}^{LR}
+\left(a_4-\frac{a_{10}}{2}\right)A_{PP}^{LL}
+\left(a_6-\frac{a_8}{2}\right)A_{PP}^{SP}\nonumber\\
&&
+\left(C_3-\frac{C_9}{2}\right)W_{PP}^{LL}
+\left(C_5-\frac{C_7}{2}\right)W_{PP}^{LR}\nonumber\\
&&+\left(C_4-\frac{C_{10}}{2}\right)M_{K}^{LL}
+\left(C_6-\frac{C_8}{2}\right)M_{K}^{SP}\Bigg],
\end{eqnarray}
}
with $PP=K^+K^-/\pi^+\pi^-$. In these decay amplitudes, the symbols $F$, $M$, $A$, and $W$ represent the contributions from different diagram types: factorizable emission diagrams, hard-rescattering emission diagrams, $W$-boson annihilation diagrams, and $W$-boson exchange diagrams, respectively. The subscript $PP(K/\pi)$ indicates that the recoiling meson is either a $KK/\pi\pi$ pair or a single $K/\pi$ meson. Meanwhile, the superscript $LL/LR/SP$ denotes the contributions from the different operator structures: $(V-A)(V-A)$, $(V-A)(V+A)$, and $(S-P)(S+P)$, respectively. For simplicity, the full expressions for these amplitudes are not provided here but can be found in our earlier work \cite{Zou:2020atb}. For the $S$-wave resonant contribution from the $f_0(980)$ resonance, the amplitude $F_{K/\pi}^{LL}$ vanishes due to charge conjugation invariance or the conservation of the vector current for the neutral scalar structure. 

Incorporating the mixing of the $f_0(980)$ and $\sigma$ meson systems, the decay amplitude for the quasi-two-body $B \to K/\pi(f_0(980) \to) P_2P_3$ decays can be written as:
\begin{eqnarray}
\mathcal{A}[B\to K/\pi(f_0(980)\to)P_2P_3]&=&\mathcal{A}[B\to K/\pi(f_0(q\bar{q})\to)P_2P_3]\cos\theta\nonumber\\
&&\,\,+\mathcal{A}[B\to K/\pi(f_0(s\bar{s})\to )P_2P_3]\sin\theta,
\end{eqnarray}
where $\theta$ is the mixing angle. The measurement and constraints on the angle $\theta$ were extensively reviewed in Refs.~\cite{Cheng:2005nb, Fleischer:2011au}. In Ref.~\cite{LHCb:2013dkk}, the LHCb collaboration analyzed the resonant components in $\overline{B}^0 \to J/\psi\pi^+\pi^-$ decays involving the $f_0(980)$ resonance, establishing an upper limit of $|\theta|< 30^\circ$. Additionally, Ref.~\cite{Cheng:2013fba} investigated hadronic $B$-decays to scalar mesons and determined $\theta = 17^\circ$, a value that yields branching fractions for $B \to f_0(980) K$ in good agreement with experimental data. In light of this consistency, we adopt $\theta = 17^\circ$ from ~\cite{Cheng:2013fba} for our numerical calculations.
\section{Numerical Results and Discussions}\label{sec:result}
We begin this section by introducing the parameters used in our numerical calculations, including the masses of the initial and final state mesons, the decay constants, the lifetimes of the $B_{(s)}$ mesons, the QCD scale, and the CKM matrix elements. The values of these parameters are summarized below \cite{ParticleDataGroup:2024cfk},
\begin{eqnarray}
&&m_{B}=5.279\,{\rm GeV},\;\; m_{B_s}=5.366\,{\rm GeV},\;\; m_{\pi^+/\pi^0}=139/135\,{\rm MeV},\;\;m_{K^+/K^0}=494/497\,{\rm MeV},\nonumber\\
&&\tau_{B^+/B^0/B_s}=1.638/1.519/1.520\,{\rm ps},\;\;V_{ud}=0.97435\pm0.00016,\;\;V_{us}=0.22500\pm0.00067,\nonumber\\
&&V_{ub}=0.00369\pm0.00011,\;\;V_{td}=0.00857^{+0.00020}_{-0.00018},\;\;V_{ts}=0.04110^{+0.00083}_{-0.00072},\nonumber\\
&&V_{tb}=0.999118^{+0.000031}_{-0.000036},\;\;\Lambda_{QCD}^{f=4}=0.25\,{\rm GeV},\;\;f_{B_{(s)}}=0.19/0.23\,{\rm GeV}.
\end{eqnarray}

Using the analytical decay amplitudes presented in Sec.~\ref{sec:function} and the input parameters, we calculate the $CP$-averaged branching fractions and direct $CP$ asymmetries of the $B \to P(f_0(980) \to) (K^+K^-,\pi^+\pi^-)$ decays, which are summarized in Table.~\ref{tb}. To assess the theoretical uncertainties, we consider three types of errors in our calculations. The first type arises from the nonperturbative parameters of the initial and final state mesons, such as the shape parameters and Gegenbauer moments in the wave functions of the $B$, $K$, $\pi$, $KK$ pair, and $\pi\pi$ pair. This also includes the decay constants, $f_B = 0.19 \pm 0.02~{\rm GeV}$ for the $B$ meson and $f_{B_s} = 0.23 \pm 0.02~{\rm GeV}$ for the $B_s$ meson. The second type of errors is estimated by varying the factorization scale $t$ from $0.8t$ to $1.2t$ and adjusting the $\Lambda_{QCD}$ from $0.2~{\rm GeV}$ to $0.3~{\rm GeV}$ to account for uncertainties due to higher-order QCD radiative corrections and higher power corrections. Finally, we consider the uncertainties associated with the CKM matrix elements. The results show that the first type of uncertainty dominates the uncertainties in the branching fractions, as the LCDAs are the primary inputs in the PQCD approach. However, this uncertainty is expected to decrease as experimental measurements improve and theoretical understanding advances. In contrast, for the direct $CP$ asymmetries, the second type of error is more significant, as higher-order corrections directly affect the strong phase. Since the uncertainties from the wave functions largely cancel out in the calculation of the direct $CP$ asymmetry, they have a reduced impact on these results.

\begin{sidewaystable}
\centering
\caption{The flavour-averaged branching ratios (in $10^{-6}$) and the localed direct $CP$ asymmetries (in $\%$) of the $B_{(s)}\to P(f_0(980)\to)KK/\pi\pi$ decays  in PQCD approach, together with the experimental data\cite{ParticleDataGroup:2022pth, BaBar:2012iuj, BaBar:2008lpx, BaBar:2009jov, Belle:2006ljg, Belle:2005rpz, Belle:2004drb}.For comparison we also list the predictions from a model based on the factorization approach (MFA)\cite{Cheng:2016shb}}\label{tb}
\begin{tabular}{l c c c c}
 \hline \hline
 \multicolumn{1}{c}{Decay Modes}&\multicolumn{1}{c}{PQCD } \;&\multicolumn{1}{c}{EXP} \;&\multicolumn{1}{c}{MFA}&\multicolumn{1}{c}{$A_{CP}^{dir}$}\\
\hline\hline
 $B^0 \to K^0(K^+K^-)$   &$8.22^{+3.68+2.56+0.14}_{-2.95-2.45-0.38}$  &$7.0^{+2.6}_{-1.8}\pm2.4$&$5.8^{+0.0+0.1+0.0}_{-0.0-0.5-0.0}$&$0.03_{-1.65-0.26-0.68}^{+1.29+0.45+0.97}$\\
$B^0 \to K^0(\pi^+\pi^-)$&$8.77^{+6.85+3.23+1.50}_{-5.76-2.83-0.00}$&$8.1\pm0.8$&$6.0_{-0.0-1.2-0.0}^{+0.0+1.5+0.0}$&$-7.31_{-3.02-0.74-0.00}^{+30.1+11.1+12.2}$\\
$B^0 \to \pi^0(K^+K^-)$&$0.053_{-0.021-0.011-0.003}^{+0.033+0.018+0.003}$&..&...&$-4.13_{-8.64-9.00-4.11}^{+9.14+7.10+3.40}$\\
$B^0 \to \pi^0(\pi^+\pi^-)$&$0.083_{-0.080-0.067-0.005}^{+0.108+0.037+0.006}$&...&...&$-77.7_{-15.0-13.6-1.5}^{+40.3+7.7+2.1}$\\
$B^+ \to K^+(K^+K^-)$&$8.50_{-3.05-2.34-0.42}^{+3.83+2.51+0.23}$&$9.4\pm1.6\pm2.8$&$11.2^{+0.0+2.7+0.0}_{-0.0-2.1-0.0}$&$-3.89_{-1.49-2.56-0.03}^{+1.24+0.74+0.64}$\\
$B^+ \to K^+(\pi^+\pi^-)$&$10.51_{-6.59-3.71-0.10}^{+8.89+4.36+0.38}$&$9.4^{+1.0}_{-1.2}$&$6.7_{-0.0-1.3-0.0}^{+0.0+1.6+0.0}$&$-8.93_{-3.35-22.4-3.14}^{+19.4+8.32+0.00}$\\
$B^+ \to \pi^+(K^+K^-)$&$0.23_{-0.08-0.03-0.02}^{+0.09+0.03+0.01}$&...&$0.25_{-0.00-0.01-0.00}^{+0.00+0.01+0.00}$&$70.5_{-6.92-1.41-0.00}^{+13.8+4.25+3.34}$\\
$B^+ \to \pi^+(\pi^+\pi^-)$&$0.63_{-0.20-0.02-0.00}^{+0.41+0.13+0.10}$&$<1.5$&$0.2_{-0.0-0.0-0.0}^{+0.0+0.0+0.0}$&$83.5_{-20.1-18.2-8.31}^{+15.1+4.86+0.41}$\\
$B_s \to \overline{K}^0(K^+K^-)$&$0.16_{-0.07-0.03-0.00}^{+0.08+0.04+0.00}$&...&...&$11.9_{-6.65-0.00-0.04}^{+17.8+8.17+3.55}$\\
$B_s \to \overline{K}^0(\pi^+\pi^-)$&$0.64_{-0.25-0.15-0.00}^{+0.42+0.23+0.02}$&...&...&$-53.0_{-5.73-0.00-0.00}^{+47.1+21.9+8.12}$\\
$B_s \to \pi^0(K^+K^-)$&$0.023_{-0.007-0.003-0.000}^{+0.011+0.003+0.000}$&...&...&$-0.21_{-3.51-10.5-0.42}^{+2.07+4.68+0.00}$\\
$B_s \to \pi^0(\pi^+\pi^-)$&$0.041_{-0.023-0.007-0.002}^{+0.029+0.012+0.001}$&...&...&$-84.3_{-15.7-15.7-5.7}^{+53.6+31.4+5.3}$\\
 \hline \hline
\end{tabular}
\end{sidewaystable}

From Table \ref{tb}, it is evident that the branching fractions for $B\to K(f_0(980)\to)(K^+K^-,\pi^+\pi^-)$ are significantly larger than those for $B\to \pi(f_0(980) \to)(K^+K^-,\pi^+\pi^-)$, which were first measured with high precision by the BaBar and Belle experiments. This difference can be understood by examining the branching fractions for the corresponding two-body decays. The branching fractions for $B\to K f_0(980)$ are notably larger than those for $B\to \pi f_0(980)$ in both the QCD factorization (QCDF) \cite{Cheng:2013fba} and PQCD approaches \cite{Wang:2006ria, Zhang:2008sa}. For instance, considering the decays $B^0\to K^0f_0(980)$ and $B^0 \to \pi^0 f_0(980)$, the branching fractions in both the QCDF and PQCD approaches are as follows:
\begin{eqnarray}
BF[B^0\to K^0f_0(980)]&=&(14.8_{-1.6-1.6-10.2}^{+1.7+1.1+28.6})\times10^{-6} ,\;\;   (QCDF)\nonumber\\
BF[B^0\to \pi^0f_0(980)]&=&(0.08_{-0.01-0.01-0.03}^{+0.01+0.01+0.08})\times10^{-6},\;\;   (QCDF)\nonumber\\
BF[B^0\to K^0f_0(980)]&=&(13\thicksim16)\times10^{-6},\;\;(PQCD)\nonumber\\
BF[B^0\to \pi^0f_0(980)]&=&(0.26\pm0.06)\times10^{-6}.\;\;(PQCD)\nonumber
\end{eqnarray}

In addition, we also find that the branching fractions for $B^{+,0} \to K^{+,0}(f_0(980) \to) (K^+K^-,\pi^+\pi^-)$ decays are larger than those for $B_s \to \overline{K}^0(f_0(980) \to) (K^+K^-,\pi^+\pi^-)$, primarily due to the suppression of the $B_s$ decays by the CKM element $V_{td}$, in contrast to the enhancement of $B^{+,0})$ decays by $V_{ts}$. The branching fractions for $B_s \to \pi^0(f_0(980) \to) (K^+K^-,\pi^+\pi^-)$ are similar to those of $B^0 \to \pi^0(f_0(980) \to) (K^+K^-,\pi^+\pi^-)$, but smaller than the enhanced $B^+ \to \pi^+(f_0(980) \to) (K^+K^-,\pi^+\pi^-)$ decays, which benefit from the isospin factor $\sqrt{2}$ and larger color-allowed tree contributions. For the $B_{u,d,s} \to \pi(f_0(980) \to) \pi^+\pi^-$ decays, the branching fractions are too small to be measurable, although they exhibit large local direct $CP$ asymmetries. Therefore, we suggest experiments focus on measuring these direct $CP$ asymmetries. In contrast, the $B \to K(f_0(980) \to) (K^+K^-,\pi^+\pi^-)$ decays have very small direct $CP$ asymmetries due to the suppression of tree operator contributions by the CKM elements $V_{ub}$ and $V_{us}$, as well as the mixing angle $\theta = 17^\circ$ of the $f_0(980)$. Since the direct $CP$ asymmetry depends on the interference between tree and penguin contributions, the suppression results in a very small asymmetry. For the $B_s \to \overline{K}^0(f_0(980) \to) \pi^+\pi^-$ decays, the central value of the direct $CP$ asymmetry is around $-53.0\%$, and the uncertainty is remarkable. We emphasize again that the theoretical description of three-body $B$ decays is still at the modeling stage, with the two-meson wave functions remaining model-dependent and the QCD-based wave function not yet available. Consequently, the large uncertainties introduced by the current two-meson wave functions affect the precision of the theoretical calculations.

Under the narrow-width approximation (NWA), the relation between the three-body resonant $B$ decays and the corresponding two-body decays is given by:
\begin{eqnarray}
BF[B\to P_1R\to P_1P_2P_3]\approx BF[B\to P_1R]\times BF[R\to P_2P_3],   
\end{eqnarray}
where $R$ represents the intermediate resonance. Based on this approximation and the theoretical predictions for the three-body $B \to K/\pi(f_0(980) \to) \pi^+ \pi^-$ decays, we can roughly estimate the branching fractions of the corresponding two-body $B \to K(\pi) f_0(980)$ decays. To perform this estimation, we need the branching fraction for $f_0(980) \to \pi^+ \pi^-$ as an input. Since there are no precise measurements for the branching fraction of $f_0(980)$ decaying into $\pi^+ \pi^-$, we follow the approach suggested in Ref.\cite{Cheng:2013fba} and use the BES measurements \cite{BES:2005iaq} on the ratio:
\begin{eqnarray}
\frac{\Gamma(f_0(980)\to \pi\pi)}{\Gamma(f_0(980)\to\pi\pi)+\Gamma(f_0(980)\to K\overline{K})}=0.75_{-0.13}^{+0.11}.
\end{eqnarray}
Assuming that the decays of $f_0(980)$ are predominantly into $\pi \pi$ and $K \overline{K}$ states, and applying isospin symmetry, the branching fraction is determined as \cite{Cheng:2013fba}:
\begin{eqnarray}
BF[f_0(980)\to\pi^+\pi^-]=0.50_{-0.09}^{+0.07},
\end{eqnarray}
which is in good agreement with the LHCb measurement of $BF[f_0(980) \to \pi^+ \pi^-] = 0.46 \pm 0.06$ from Ref.\cite{LHCb:2013dkk}. With this, we can estimate the branching fractions for the $B \to K(\pi) f_0(980)$ decays using the narrow-width approximation:
\begin{eqnarray}
BF[B^0\to K^0f_0(980)]&\thickapprox&\frac{BF[B^0 \to K^0(f_0(980)\to)\pi^+\pi^-]}{BF[f_0(980)\to \pi^+\pi^-]}\thickapprox (17.54_{-11.67}^{+15.36})\times  10^{-6}\nonumber\\
BF[B^0\to \pi^0f_0(980)]&\thickapprox&\frac{BF[B^0 \to \pi^0 (f_0(980)\to)\pi^+\pi^-]}{BF[f_0(980)\to \pi^+\pi^-]}\thickapprox (0.17_{-0.15}^{+0.23})\times 10^{-6}\nonumber\\
BF[B^+\to K^+f_0(980)]&\thickapprox&\frac{BF[B^+ \to K^+(f_0(980)\to)\pi^+\pi^-]}{BF[f_0(980)\to \pi^+\pi^-]}\thickapprox(21.02_{-15.10}^{+19.80})\times 10^{-6}\nonumber\\
BF[B^+\to \pi^+f_0(980)]&\thickapprox&\frac{BF[B^+ \to \pi^+(f_0(980)\to)\pi^+\pi^-]}{BF[f_0(980)\to \pi^+\pi^-]}\thickapprox(1.26_{-0.40}^{+0.88})\times 10^{-6}\nonumber\\
BF[B_s\to \overline{K}^0 f_0(980)]&\thickapprox&\frac{BF[B_s \to \overline{K}^0(f_0(980)\to)\pi^+\pi^-]}{BF[f_0(980)\to \pi^+\pi^-]}\thickapprox(1.28_{-0.58}^{+0.94})\times 10^{-6}\nonumber\\
BF[B_s\to \pi^0f_0(980)]&\thickapprox&\frac{BF[B_s \to \pi^0(f_0(980)\to)\pi^+\pi^-]}{BF[f_0(980)\to \pi^+\pi^-]}\thickapprox(0.082_{-0.048}^{+0.062})\times 10^{-6}.
\end{eqnarray}
These results are consistent with the predictions from both QCDF \cite{Cheng:2013fba} and the previous PQCD approach \cite{Wang:2006ria, Zhang:2008sa}. They are expected to be experimentally measurable in the near future.

It is crucial to emphasize that we did not use the $B \to P(f_0(980) \to) K^+K^-$ decays to extract the branching fractions of the two-body $B \to P f_0(980)$ decays. This issue was addressed in our previous study on $B \to KKK$ decays \cite{Zou:2020atb}, which aligns with the discussion in Ref.~\cite{Cheng:2007si}. We explained that NWA is valid only when the decays $B \to P_1 R$ and $R \to P_2 P_3$ are kinematically allowed, and when the resonance width is small. However, in the case of $f_0(980) \to K^+K^-$, the kinematics do not permit this decay, or it is effectively forbidden. As a result, in the three-body decay $B \to P(f_0(980) \to) K^+K^-$, the width of the resonance cannot be neglected. This means that the intermediate $f_0(980)$ is off-shell, and we cannot apply the NWA to factorize the decay rate, i.e., $\Gamma[B \to P f_0(980) \to P K^+K^-] \neq \Gamma[B \to P f_0(980)] \times \Gamma[f_0(980) \to K \overline{K}]$. Therefore, using NWA to extract the branching fractions of the corresponding two-body decays is not feasible. In contrast, since $f_0(980) \to \pi^+\pi^-$ is kinematically allowed, it is more appropriate to use decays like $B \to P \pi^+\pi^-$ rather than $B \to P K^+K^-$ to extract the branching fractions for $B \to P f_0(980)$ decays, assuming the branching ratio $BF[f_0(980) \to \pi^+\pi^-]$ is known.

\section{Summary}\label{sec:summary}
In this work, we calculate the quasi-two-body decays  $B\to P f_0(980)\to P(K^+K^-, \pi^+\pi^-)$ , where $P = K, \pi$, within the PQCD approach. We apply the updated relativistic Breit-Wigner formula for the $S$-wave resonance $f_0(980)$ to parameterize the timelike form factors $F_s(\omega)$, which encapsulate the final-state interactions between the collinear mesons in the resonant regions. Using the Gegenbauer moments of the  $S$-wave meson pair distribution amplitudes, we predict the branching fractions and direct $CP$  asymmetries for the $B\to P f_0(980)\to P (K^+K^-, \pi^+\pi^-)$ channels, and compare the differential branching fractions with available experimental data. While data is currently limited, the PQCD predictions generally agree with the existing measurements. Under the narrow-width approximation, we extract the branching fractions of the corresponding two-body decays  $B\to Pf_0(980)$ from the quasi-two-body decay modes $B \to P (f_0(980) \to) \pi^+\pi^-$, with the related branching fractions being consistent with previous studies and experimental data. We also note that the narrow-width approximation is not valid for extracting branching fractions from the quasi-two-body decay  $B \to P f_0(980) \to P K^+K^-$ due to kinematical reasons. More precise data from LHCb and future Belle II experiments will provide further tests of our predictions.
\section*{Acknowledgment}
This work was supported by National Natural Science Foundation of China under Grants Nos.~12375089 and 12435004, and by the Natural Science Foundation of Shandong province under the Grant No. 
\bibliographystyle{bibstyle}
\bibliography{refs}

\providecommand{\href}[2]{#2}\begingroup\raggedright\begin{thebibliography}{10}

\bibitem{Kobayashi:1973fv}
M.~Kobayashi and T.~Maskawa, {\it {CP Violation in the Renormalizable Theory of
  Weak Interaction}},  {\em Prog. Theor. Phys.} {\bf 49} (1973) 652--657.

\bibitem{BaBar:2014omp}
{\bf BaBar, Belle} Collaboration, A.~J. Bevan et~al., {\it {The Physics of the
  $B$ Factories}},  {\em Eur. Phys. J. C} {\bf 74} (2014) 3026,
  [\href{https://arxiv.org/abs/1406.6311}{{\tt arXiv:1406.6311}}].

\bibitem{Kuhr:2013hd}
T.~Kuhr, {\it {Flavor physics at the Tevatron}},  {\em Springer Tracts
  Mod.Phys.} {\bf 249} (2013) 1--161.

\bibitem{LHCb:2012myk}
{\bf LHCb} Collaboration, R.~Aaij et~al., {\it {Implications of LHCb
  measurements and future prospects}},  {\em Eur. Phys. J. C} {\bf 73} (2013),
  no.~4 2373, [\href{https://arxiv.org/abs/1208.3355}{{\tt arXiv:1208.3355}}].

\bibitem{Belle-II:2018jsg}
{\bf Belle-II} Collaboration, W.~Altmannshofer et~al., {\it {The Belle II
  Physics Book}},  {\em PTEP} {\bf 2019} (2019), no.~12 123C01,
  [\href{https://arxiv.org/abs/1808.10567}{{\tt arXiv:1808.10567}}]. [Erratum:
  PTEP 2020, 029201 (2020)].

\bibitem{Virto:2016fbw}
J.~Virto, {\it {Charmless Non-Leptonic Multi-Body B decays}},  {\em PoS} {\bf
  FPCP2016} (2017) 007, [\href{https://arxiv.org/abs/1609.07430}{{\tt
  arXiv:1609.07430}}].

\bibitem{Liu:2021sdw}
W.-F. Liu, Z.-T. Zou, and Y.~Li, {\it {Charmless Quasi-Two-Body B Decays in
  Perturbative QCD Approach: Taking $B\to K(R\to K^+K^-)$ as Examples}},  {\em
  Adv. High Energy Phys.} {\bf 2022} (2022) 5287693,
  [\href{https://arxiv.org/abs/2112.00315}{{\tt arXiv:2112.00315}}].

\bibitem{Dalitz:1953cp}
R.~H. Dalitz, {\it {On the analysis of tau-meson data and the nature of the
  tau-meson}},  {\em Phil. Mag. Ser. 7} {\bf 44} (1953) 1068--1080.

\bibitem{El-Bennich:2009gqk}
B.~El-Bennich, A.~Furman, R.~Kaminski, L.~Lesniak, B.~Loiseau, and
  B.~Moussallam, {\it {CP violation and kaon-pion interactions in $B \to K
  \pi^+ \pi^-$ decays}},  {\em Phys. Rev. D} {\bf 79} (2009) 094005,
  [\href{https://arxiv.org/abs/0902.3645}{{\tt arXiv:0902.3645}}]. [Erratum:
  Phys.Rev.D 83, 039903 (2011)].

\bibitem{Krankl:2015fha}
S.~Kr\"ankl, T.~Mannel, and J.~Virto, {\it {Three-body non-leptonic B decays
  and QCD factorization}},  {\em Nucl. Phys. B} {\bf 899} (2015) 247--264,
  [\href{https://arxiv.org/abs/1505.04111}{{\tt arXiv:1505.04111}}].

\bibitem{Klein:2017xti}
R.~Klein, T.~Mannel, J.~Virto, and K.~K. Vos, {\it {CP Violation in Multibody
  $B$ Decays from QCD Factorization}},  {\em JHEP} {\bf 10} (2017) 117,
  [\href{https://arxiv.org/abs/1708.02047}{{\tt arXiv:1708.02047}}].

\bibitem{Cheng:2016shb}
H.-Y. Cheng, C.-K. Chua, and Z.-Q. Zhang, {\it {Direct CP Violation in
  Charmless Three-body Decays of $B$ Mesons}},  {\em Phys. Rev. D} {\bf 94}
  (2016), no.~9 094015, [\href{https://arxiv.org/abs/1607.08313}{{\tt
  arXiv:1607.08313}}].

\bibitem{Cheng:2014uga}
H.-Y. Cheng and C.-K. Chua, {\it {Charmless three-body decays of $B_s$
  mesons}},  {\em Phys. Rev. D} {\bf 89} (2014), no.~7 074025,
  [\href{https://arxiv.org/abs/1401.5514}{{\tt arXiv:1401.5514}}].

\bibitem{Li:2014oca}
Y.~Li, {\it {Comprehensive study of $\overline B^0\to K^0(\overline K^0)
  K^\mp\pi^\pm$ decays in the factorization approach}},  {\em Phys. Rev. D}
  {\bf 89} (2014), no.~9 094007, [\href{https://arxiv.org/abs/1402.6052}{{\tt
  arXiv:1402.6052}}].

\bibitem{Lesniak:2024zrc}
L.~Lesniak and P.~Zenczykowski, {\it {Dalitz-plot analysis of $B^{\pm}\to K^\pm
  K^+K^-$ decays}},  {\em Phys. Rev. D} {\bf 110} (2024), no.~3 033001,
  [\href{https://arxiv.org/abs/2405.18192}{{\tt arXiv:2405.18192}}].

\bibitem{Wang:2016rlo}
W.-F. Wang and H.-n. Li, {\it {Quasi-two-body decays $B\to K\rho\to K\pi\pi$ in
  perturbative QCD approach}},  {\em Phys. Lett. B} {\bf 763} (2016) 29--39,
  [\href{https://arxiv.org/abs/1609.04614}{{\tt arXiv:1609.04614}}].

\bibitem{Li:2016tpn}
Y.~Li, A.-J. Ma, W.-F. Wang, and Z.-J. Xiao, {\it {Quasi-two-body decays
  $B_{(s)}\to P\rho\to P\pi\pi$ in perturbative QCD approach}},  {\em Phys.
  Rev. D} {\bf 95} (2017), no.~5 056008,
  [\href{https://arxiv.org/abs/1612.05934}{{\tt arXiv:1612.05934}}].

\bibitem{Rui:2017bgg}
Z.~Rui, Y.~Li, and W.-F. Wang, {\it {The S-wave resonance contributions in the
  $B^0_s$ decays into $ \psi(2S,3S)$ plus pion pair}},  {\em Eur. Phys. J. C}
  {\bf 77} (2017), no.~3 199, [\href{https://arxiv.org/abs/1701.02941}{{\tt
  arXiv:1701.02941}}].

\bibitem{Zou:2020atb}
Z.-T. Zou, Y.~Li, Q.-X. Li, and X.~Liu, {\it {Resonant contributions to
  three-body $B\rightarrow KKK$ decays in perturbative QCD approach}},  {\em
  Eur. Phys. J. C} {\bf 80} (2020), no.~5 394,
  [\href{https://arxiv.org/abs/2003.03754}{{\tt arXiv:2003.03754}}].

\bibitem{Zou:2020fax}
Z.-T. Zou, Y.~Li, and X.~Liu, {\it {Branching fractions and CP asymmetries of
  the quasi-two-body decays in $B_{s} \rightarrow K^0({\overline{K}}^0)K^\pm
  \pi ^\mp $ within PQCD approach}},  {\em Eur. Phys. J. C} {\bf 80} (2020),
  no.~6 517, [\href{https://arxiv.org/abs/2005.02097}{{\tt arXiv:2005.02097}}].

\bibitem{Zou:2020ool}
Z.-T. Zou, L.~Yang, Y.~Li, and X.~Liu, {\it {Study of Quasi-two-body
  $B_{(s)}\to \phi (f_0(980)/f_2(1270)\to)\pi\pi$ Decays in Perturbative QCD
  Approach}},  {\em Eur. Phys. J. C} {\bf 81} (2021), no.~1 91,
  [\href{https://arxiv.org/abs/2011.07676}{{\tt arXiv:2011.07676}}].

\bibitem{Zhang:2013oqa}
Z.-H. Zhang, X.-H. Guo, and Y.-D. Yang, {\it {CP violation in $B^{\pm}
  \rightarrow \pi^{\pm}\pi^{+}\pi^{-}$ in the region with low invariant mass of
  one $\pi^{+}\pi^{-}$ pair}},  {\em Phys. Rev. D} {\bf 87} (2013), no.~7
  076007, [\href{https://arxiv.org/abs/1303.3676}{{\tt arXiv:1303.3676}}].

\bibitem{El-Bennich:2006rcn}
B.~El-Bennich, A.~Furman, R.~Kaminski, L.~Lesniak, and B.~Loiseau, {\it
  {Interference between $f_0(980)$ and $\rho(770)^0$ resonances in $B\to \pi^+
  \pi^- K$ decays}},  {\em Phys. Rev. D} {\bf 74} (2006) 114009,
  [\href{https://arxiv.org/abs/hep-ph/0608205}{{\tt hep-ph/0608205}}].

\bibitem{Hu:2022eql}
R.~Hu and Z.-H. Zhang, {\it {Data-based analysis of the forward-backward
  asymmetry in $B^\pm\to K^\pm K^\mp K^\pm$}},  {\em Phys. Rev. D} {\bf 105}
  (2022), no.~9 093007, [\href{https://arxiv.org/abs/2201.07456}{{\tt
  arXiv:2201.07456}}].

\bibitem{Abreu:2023hts}
L.~M. Abreu, N.~Ikeno, and E.~Oset, {\it {Role of $f_0(980)$ and $a_0(980)$ in
  the $B^-\to {\pi}^-K^+K^-$ and $B^-\to {\pi}^-K^0\bar K^0$ reactions}},  {\em
  Phys. Rev. D} {\bf 108} (2023), no.~1 016007,
  [\href{https://arxiv.org/abs/2305.02848}{{\tt arXiv:2305.02848}}].

\bibitem{ParticleDataGroup:2024cfk}
{\bf Particle Data Group} Collaboration, S.~Navas et~al., {\it {Review of
  particle physics}},  {\em Phys. Rev. D} {\bf 110} (2024), no.~3 030001.

\bibitem{Achasov:2020fee}
N.~N. Achasov, {\it {Fate of Light Scalar Mesons}},  {\em Phys. Part. Nucl.}
  {\bf 51} (2020), no.~4 632--639,
  [\href{https://arxiv.org/abs/2002.01354}{{\tt arXiv:2002.01354}}].

\bibitem{Close:2002zu}
F.~E. Close and N.~A. Tornqvist, {\it {Scalar mesons above and below 1-GeV}},
  {\em J. Phys. G} {\bf 28} (2002) R249--R267,
  [\href{https://arxiv.org/abs/hep-ph/0204205}{{\tt hep-ph/0204205}}].

\bibitem{Weinstein:1990gu}
J.~D. Weinstein and N.~Isgur, {\it {K anti-K Molecules}},  {\em Phys. Rev. D}
  {\bf 41} (1990) 2236.

\bibitem{Achasov:2020aun}
N.~N. Achasov, J.~V. Bennett, A.~V. Kiselev, E.~A. Kozyrev, and G.~N.
  Shestakov, {\it {Evidence of the four-quark nature of $f_0$(980) and
  $f_0$(500)}},  {\em Phys. Rev. D} {\bf 103} (2021), no.~1 014010,
  [\href{https://arxiv.org/abs/2009.04191}{{\tt arXiv:2009.04191}}].

\bibitem{BaBar:2005qms}
{\bf BaBar} Collaboration, B.~Aubert et~al., {\it {Dalitz-plot analysis of the
  decays $B^\pm \to K^\pm \pi^\mp \pi^\pm$}},  {\em Phys. Rev. D} {\bf 72}
  (2005) 072003, [\href{https://arxiv.org/abs/hep-ex/0507004}{{\tt
  hep-ex/0507004}}]. [Erratum: Phys.Rev.D 74, 099903 (2006)].

\bibitem{BaBar:2007itz}
{\bf BaBar} Collaboration, B.~Aubert et~al., {\it {Observation of the Decay
  $B^{+} \to K^{+} K^{-} \pi^{+}$}},  {\em Phys. Rev. Lett.} {\bf 99} (2007)
  221801, [\href{https://arxiv.org/abs/0708.0376}{{\tt arXiv:0708.0376}}].

\bibitem{BaBar:2008lpx}
{\bf BaBar} Collaboration, B.~Aubert et~al., {\it {Evidence for Direct CP
  Violation from Dalitz-plot analysis of $B^\pm \to K^\pm \pi^\mp \pi^\pm$}},
  {\em Phys. Rev. D} {\bf 78} (2008) 012004,
  [\href{https://arxiv.org/abs/0803.4451}{{\tt arXiv:0803.4451}}].

\bibitem{BaBar:2009jov}
{\bf BaBar} Collaboration, B.~Aubert et~al., {\it {Time-dependent amplitude
  analysis of $B^0\to K_s^0 \pi^+ \pi^-$}},  {\em Phys. Rev. D} {\bf 80} (2009)
  112001, [\href{https://arxiv.org/abs/0905.3615}{{\tt arXiv:0905.3615}}].

\bibitem{BaBar:2009vfr}
{\bf BaBar} Collaboration, B.~Aubert et~al., {\it {Dalitz Plot Analysis of
  $B^\pm\to \pi^\pm \pi^\pm \pi^\mp$ Decays}},  {\em Phys. Rev. D} {\bf 79}
  (2009) 072006, [\href{https://arxiv.org/abs/0902.2051}{{\tt
  arXiv:0902.2051}}].

\bibitem{BaBar:2011ktx}
{\bf BaBar} Collaboration, J.~P. Lees et~al., {\it {Amplitude analysis and
  measurement of the time-dependent CP asymmetry of $B^0 \to K_S^0 K_S^0 K_S^0$
  decays}},  {\em Phys. Rev. D} {\bf 85} (2012) 054023,
  [\href{https://arxiv.org/abs/1111.3636}{{\tt arXiv:1111.3636}}].

\bibitem{BaBar:2011vfx}
{\bf BaBar} Collaboration, J.~P. Lees et~al., {\it {Amplitude Analysis of
  $B^0\to K^+ \pi^- \pi^0$ and Evidence of Direct CP Violation in $B\to K^*
  \pi$ decays}},  {\em Phys. Rev. D} {\bf 83} (2011) 112010,
  [\href{https://arxiv.org/abs/1105.0125}{{\tt arXiv:1105.0125}}].

\bibitem{BaBar:2012iuj}
{\bf BaBar} Collaboration, J.~P. Lees et~al., {\it {Study of CP violation in
  Dalitz-plot analyses of $B^0\to K^+K^-K_s^0$,$B^+\to K^+K^-K^+$, and $B^+\to
  K_s^0K_s^0K^+$}},  {\em Phys. Rev. D} {\bf 85} (2012) 112010,
  [\href{https://arxiv.org/abs/1201.5897}{{\tt arXiv:1201.5897}}].

\bibitem{Belle:2004drb}
{\bf Belle} Collaboration, A.~Garmash et~al., {\it {Dalitz analysis of the
  three-body charmless decays $B^+ \to K^+ \pi^+ \pi^-$ and $B^+ \to K^+ K^+
  K^-$ }},  {\em Phys. Rev. D} {\bf 71} (2005) 092003,
  [\href{https://arxiv.org/abs/hep-ex/0412066}{{\tt hep-ex/0412066}}].

\bibitem{Belle:2005rpz}
{\bf Belle} Collaboration, A.~Garmash et~al., {\it {Evidence for large direct
  CP violation in $B^\pm\to \rho(770)^0 K^\pm$ from analysis of the three-body
  charmless $B^\pm \to K^\pm \pi^\pm \pi^\mp$ decay}},  {\em Phys. Rev. Lett.}
  {\bf 96} (2006) 251803, [\href{https://arxiv.org/abs/hep-ex/0512066}{{\tt
  hep-ex/0512066}}].

\bibitem{Belle:2006ljg}
{\bf Belle} Collaboration, A.~Garmash et~al., {\it {Dalitz Analysis of
  Three-body Charmless $B^0\to K^0 \pi^+ \pi^-$ Decay}},  {\em Phys. Rev. D}
  {\bf 75} (2007) 012006, [\href{https://arxiv.org/abs/hep-ex/0610081}{{\tt
  hep-ex/0610081}}].

\bibitem{Belle:2008til}
{\bf Belle} Collaboration, J.~Dalseno et~al., {\it {Time-dependent Dalitz Plot
  Measurement of CP Parameters in $B^0 \to K^0_S \pi^+ \pi^-$ Decays}},  {\em
  Phys. Rev. D} {\bf 79} (2009) 072004,
  [\href{https://arxiv.org/abs/0811.3665}{{\tt arXiv:0811.3665}}].

\bibitem{Belle:2010wis}
{\bf Belle} Collaboration, Y.~Nakahama et~al., {\it {Measurement of CP
  violating asymmetries in $B^0 \to K^+K^- K^0_S$ decays with a time-dependent
  Dalitz approach}},  {\em Phys. Rev. D} {\bf 82} (2010) 073011,
  [\href{https://arxiv.org/abs/1007.3848}{{\tt arXiv:1007.3848}}].

\bibitem{Belle:2017cxf}
{\bf Belle} Collaboration, C.~L. Hsu et~al., {\it {Measurement of branching
  fraction and direct $CP$ asymmetry in charmless $B^+ \to K^+K^- \pi^+$ decays
  at Belle}},  {\em Phys. Rev. D} {\bf 96} (2017), no.~3 031101,
  [\href{https://arxiv.org/abs/1705.02640}{{\tt arXiv:1705.02640}}].

\bibitem{LHCb:2013dkk}
{\bf LHCb} Collaboration, R.~Aaij et~al., {\it {Analysis of the resonant
  components in $\bar B^0 \to J/\psi\pi^+\pi^-$}},  {\em Phys. Rev. D} {\bf 87}
  (2013), no.~5 052001, [\href{https://arxiv.org/abs/1301.5347}{{\tt
  arXiv:1301.5347}}].

\bibitem{LHCb:2013lcl}
{\bf LHCb} Collaboration, R.~Aaij et~al., {\it {Measurement of CP violation in
  the phase space of $B^{\pm} \rightarrow K^{+} K^{-} \pi^{\pm}$ and $B^{\pm}
  \rightarrow \pi^{+} \pi^{-} \pi^{\pm}$ decays}},  {\em Phys. Rev. Lett.} {\bf
  112} (2014), no.~1 011801, [\href{https://arxiv.org/abs/1310.4740}{{\tt
  arXiv:1310.4740}}].

\bibitem{LHCb:2013ptu}
{\bf LHCb} Collaboration, R.~Aaij et~al., {\it {Measurement of CP violation in
  the phase space of $B^{\pm} \to K^{\pm} \pi^{+} \pi^{-}$ and $B^{\pm} \to
  K^{\pm} K^{+} K^{-}$ decays}},  {\em Phys. Rev. Lett.} {\bf 111} (2013)
  101801, [\href{https://arxiv.org/abs/1306.1246}{{\tt arXiv:1306.1246}}].

\bibitem{LHCb:2014mir}
{\bf LHCb} Collaboration, R.~Aaij et~al., {\it {Measurements of $CP$ violation
  in the three-body phase space of charmless $B^{\pm}$ decays}},  {\em Phys.
  Rev. D} {\bf 90} (2014), no.~11 112004,
  [\href{https://arxiv.org/abs/1408.5373}{{\tt arXiv:1408.5373}}].

\bibitem{LHCb:2019jta}
{\bf LHCb} Collaboration, R.~Aaij et~al., {\it {Observation of Several Sources
  of $CP$ Violation in $B^+ \to \pi^+ \pi^+ \pi^-$ Decays}},  {\em Phys. Rev.
  Lett.} {\bf 124} (2020), no.~3 031801,
  [\href{https://arxiv.org/abs/1909.05211}{{\tt arXiv:1909.05211}}].

\bibitem{LHCb:2019sus}
{\bf LHCb} Collaboration, R.~Aaij et~al., {\it {Amplitude analysis of the $B^+
  \rightarrow \pi^+\pi^+\pi^-$ decay}},  {\em Phys. Rev. D} {\bf 101} (2020),
  no.~1 012006, [\href{https://arxiv.org/abs/1909.05212}{{\tt
  arXiv:1909.05212}}].

\bibitem{LHCb:2019vww}
{\bf LHCb} Collaboration, R.~Aaij et~al., {\it {Amplitude analysis of
  $B^{0}_{s} \rightarrow K^{0}_{\textrm{S}} K^{\pm}\pi^{\mp}$ decays}},  {\em
  JHEP} {\bf 06} (2019) 114, [\href{https://arxiv.org/abs/1902.07955}{{\tt
  arXiv:1902.07955}}].

\bibitem{LHCb:2019xmb}
{\bf LHCb} Collaboration, R.~Aaij et~al., {\it {Amplitude analysis of $B^{\pm}
  \to \pi^{\pm} K^{+} K^{-}$ decays}},  {\em Phys. Rev. Lett.} {\bf 123}
  (2019), no.~23 231802, [\href{https://arxiv.org/abs/1905.09244}{{\tt
  arXiv:1905.09244}}].

\bibitem{LHCb:2022fpg}
{\bf LHCb} Collaboration, R.~Aaij et~al., {\it {Direct CP violation in
  charmless three-body decays of B\ensuremath{\pm} mesons}},  {\em Phys. Rev.
  D} {\bf 108} (2023), no.~1 012008,
  [\href{https://arxiv.org/abs/2206.07622}{{\tt arXiv:2206.07622}}].

\bibitem{Herndon:1973yn}
D.~Herndon, P.~Soding, and R.~J. Cashmore, {\it {A GENERALIZED ISOBAR MODEL
  FORMALISM}},  {\em Phys. Rev. D} {\bf 11} (1975) 3165.

\bibitem{Chung:1995dx}
S.~U. Chung, J.~Brose, R.~Hackmann, E.~Klempt, S.~Spanier, and C.~Strassburger,
  {\it {Partial wave analysis in K matrix formalism}},  {\em Annalen Phys.}
  {\bf 4} (1995) 404--430.

\bibitem{Buchalla:1995vs}
G.~Buchalla, A.~J. Buras, and M.~E. Lautenbacher, {\it {Weak decays beyond
  leading logarithms}},  {\em Rev. Mod. Phys.} {\bf 68} (1996) 1125--1144,
  [\href{https://arxiv.org/abs/hep-ph/9512380}{{\tt hep-ph/9512380}}].

\bibitem{Lu:2000em}
C.-D. Lu, K.~Ukai, and M.-Z. Yang, {\it {Branching ratio and CP violation of $B
  \to \pi \pi$ decays in perturbative QCD approach}},  {\em Phys. Rev. D} {\bf
  63} (2001) 074009, [\href{https://arxiv.org/abs/hep-ph/0004213}{{\tt
  hep-ph/0004213}}].

\bibitem{Yu:2005rh}
X.-Q. Yu, Y.~Li, and C.-D. Lu, {\it {Branching ratio and CP violation of $B_s
  \to \pi K$ decays in the perturbative QCD approach}},  {\em Phys. Rev. D}
  {\bf 71} (2005) 074026, [\href{https://arxiv.org/abs/hep-ph/0501152}{{\tt
  hep-ph/0501152}}]. [Erratum: Phys.Rev.D 72, 119903 (2005)].

\bibitem{Ali:2007ff}
A.~Ali, G.~Kramer, Y.~Li, C.-D. Lu, Y.-L. Shen, W.~Wang, and Y.-M. Wang, {\it
  {Charmless non-leptonic $B_s$ decays to $PP$, $PV$ and $VV$ final states in
  the pQCD approach}},  {\em Phys. Rev. D} {\bf 76} (2007) 074018,
  [\href{https://arxiv.org/abs/hep-ph/0703162}{{\tt hep-ph/0703162}}].

\bibitem{Zou:2015iwa}
Z.-T. Zou, A.~Ali, C.-D. Lu, X.~Liu, and Y.~Li, {\it {Improved Estimates of The
  $B_{(s)}\to V V$ Decays in Perturbative QCD Approach}},  {\em Phys. Rev. D}
  {\bf 91} (2015) 054033, [\href{https://arxiv.org/abs/1501.00784}{{\tt
  arXiv:1501.00784}}].

\bibitem{Li:2019hnt}
Y.~Li, D.-C. Yan, Z.~Rui, and Z.-J. Xiao, {\it {$S$, $P$ and $D$-wave resonance
  contributions to $B_{(s)} \to \eta_c(1S,2S) K\pi$ decays in the perturbative
  QCD approach}},  {\em Phys. Rev. D} {\bf 101} (2020), no.~1 016015,
  [\href{https://arxiv.org/abs/1911.09348}{{\tt arXiv:1911.09348}}].

\bibitem{Rui:2019yxx}
Z.~Rui, Y.~Li, and H.~Li, {\it {Studies of the resonance components in the
  $B_s$ decays into charmonia plus kaon pair}},  {\em Eur. Phys. J. C} {\bf 79}
  (2019), no.~9 792, [\href{https://arxiv.org/abs/1907.04128}{{\tt
  arXiv:1907.04128}}].

\bibitem{Wang:2015uea}
W.-F. Wang, H.-n. Li, W.~Wang, and C.-D. L\"u, {\it {$S$-wave resonance
  contributions to the $B^0_{(s)}\to J/\psi\pi^+\pi^-$ and
  $B_s\to\pi^+\pi^-\mu^+\mu^-$ decays}},  {\em Phys. Rev. D} {\bf 91} (2015),
  no.~9 094024, [\href{https://arxiv.org/abs/1502.05483}{{\tt
  arXiv:1502.05483}}].

\bibitem{Xing:2019xti}
Y.~Xing and Z.-P. Xing, {\it {$S$-wave contributions in $\bar B_s^0\to
  (D^0,\bar D^0)\pi^+\pi^- $ within perturbative QCD approach}},  {\em Chin.
  Phys. C} {\bf 43} (2019), no.~7 073103,
  [\href{https://arxiv.org/abs/1903.04255}{{\tt arXiv:1903.04255}}].

\bibitem{Li:2020zng}
Y.~Li, D.-C. Yan, Z.~Rui, L.~Liu, Y.-T. Zhang, and Z.-J. Xiao, {\it {Resonant
  contributions to three-body $B_{(s)} \to [ D^{(*)}, \bar{D}^{(*)} ] K^+K^-$
  decays in the perturbative QCD approach}},  {\em Phys. Rev. D} {\bf 102}
  (2020), no.~5 056017, [\href{https://arxiv.org/abs/2007.13629}{{\tt
  arXiv:2007.13629}}].

\bibitem{Jia:2021uhi}
M.-K. Jia, C.-Q. Zhang, J.-M. Li, and Z.~Rui, {\it {S-wave contributions to the
  $B_{(s)} \rightarrow$ $\chi_{c1} (\pi \pi, K\pi, KK)$ decays}},  {\em Phys.
  Rev. D} {\bf 104} (2021), no.~7 073001,
  [\href{https://arxiv.org/abs/2107.13882}{{\tt arXiv:2107.13882}}].

\bibitem{ParticleDataGroup:2018ovx}
{\bf Particle Data Group} Collaboration, M.~Tanabashi et~al., {\it {Review of
  Particle Physics}},  {\em Phys. Rev. D} {\bf 98} (2018), no.~3 030001.

\bibitem{Alston-Garnjost:1971lsd}
M.~Alston-Garnjost, A.~Barbaro-Galtieri, S.~M. Flatte, J.~H. Friedman, G.~R.
  Lynch, S.~D. Protopopescu, M.~S. Rabin, and F.~T. Solmitz, {\it {OBSERVATION
  OF AN ANOMALY IN THE $\pi^+\pi^-$ SYSTEM AT 980-MeV}},  {\em Phys. Lett. B}
  {\bf 36} (1971) 152--156.

\bibitem{Flatte:1972rz}
S.~M. Flatte, M.~Alston-Garnjost, A.~Barbaro-Galtieri, J.~H. Friedman, G.~R.
  Lynch, S.~D. Protopopescu, M.~S. Rabin, and F.~T. Solmitz, {\it {ANALYSIS OF
  THE OBSERVED ANOMALY IN $\pi\pi$s WAVE SCATTERING NEAR K anti-K THRESHOLD}},
  {\em Phys. Lett. B} {\bf 38} (1972) 232--236.

\bibitem{Flatte:1976xu}
S.~M. Flatte, {\it {Coupled - Channel Analysis of the $\pi\eta$ and $K\bar K$
  Systems Near $K\bar K$ Threshold}},  {\em Phys. Lett. B} {\bf 63} (1976)
  224--227.

\bibitem{Bugg:2008ig}
D.~V. Bugg, {\it {Re-analysis of data on $a_0(1450)$ and $a_0(980)$}},  {\em
  Phys. Rev. D} {\bf 78} (2008) 074023,
  [\href{https://arxiv.org/abs/0808.2706}{{\tt arXiv:0808.2706}}].

\bibitem{LHCb:2014ooi}
{\bf LHCb} Collaboration, R.~Aaij et~al., {\it {Measurement of resonant and CP
  components in $\bar{B}_s^0\to J/\psi\pi^+\pi^-$ decays}},  {\em Phys. Rev. D}
  {\bf 89} (2014), no.~9 092006, [\href{https://arxiv.org/abs/1402.6248}{{\tt
  arXiv:1402.6248}}].

\bibitem{Cheng:2005nb}
H.-Y. Cheng, C.-K. Chua, and K.-C. Yang, {\it {Charmless hadronic B decays
  involving scalar mesons: Implications to the nature of light scalar mesons}},
   {\em Phys. Rev. D} {\bf 73} (2006) 014017,
  [\href{https://arxiv.org/abs/hep-ph/0508104}{{\tt hep-ph/0508104}}].

\bibitem{Fleischer:2011au}
R.~Fleischer, R.~Knegjens, and G.~Ricciardi, {\it {Anatomy of $B^0_{s,d} \to
  J/\psi f_0(980)$}},  {\em Eur. Phys. J. C} {\bf 71} (2011) 1832,
  [\href{https://arxiv.org/abs/1109.1112}{{\tt arXiv:1109.1112}}].

\bibitem{Cheng:2013fba}
H.-Y. Cheng, C.-K. Chua, K.-C. Yang, and Z.-Q. Zhang, {\it {Revisiting
  charmless hadronic B decays to scalar mesons}},  {\em Phys. Rev. D} {\bf 87}
  (2013), no.~11 114001, [\href{https://arxiv.org/abs/1303.4403}{{\tt
  arXiv:1303.4403}}].

\bibitem{ParticleDataGroup:2022pth}
{\bf Particle Data Group} Collaboration, R.~L. Workman et~al., {\it {Review of
  Particle Physics}},  {\em PTEP} {\bf 2022} (2022) 083C01.

\bibitem{Wang:2006ria}
W.~Wang, Y.-L. Shen, Y.~Li, and C.-D. Lu, {\it {Study of scalar mesons
  $f_0(980)$ and $f_0(1500)$ from $B \to f_0(980) K$ and $B \to f_0(980) K$
  Decays}},  {\em Phys. Rev. D} {\bf 74} (2006) 114010,
  [\href{https://arxiv.org/abs/hep-ph/0609082}{{\tt hep-ph/0609082}}].

\bibitem{Zhang:2008sa}
Z.-Q. Zhang and Z.-J. Xiao, {\it {$B \to f_0(980)(\pi, \eta^{(\prime)})$ Decays
  in the PQCD Approach}},  {\em Chin. Phys. C} {\bf 33} (2009) 508--515,
  [\href{https://arxiv.org/abs/0812.2314}{{\tt arXiv:0812.2314}}].

\bibitem{BES:2005iaq}
{\bf BES} Collaboration, M.~Ablikim et~al., {\it {Partial wave analysis of
  $\chi_{c0} \to \pi^+ \pi^- K^+ K^-$}},  {\em Phys. Rev. D} {\bf 72} (2005)
  092002, [\href{https://arxiv.org/abs/hep-ex/0508050}{{\tt hep-ex/0508050}}].

\bibitem{Cheng:2007si}
H.-Y. Cheng, C.-K. Chua, and A.~Soni, {\it {Charmless three-body decays of B
  mesons}},  {\em Phys. Rev. D} {\bf 76} (2007) 094006,
  [\href{https://arxiv.org/abs/0704.1049}{{\tt arXiv:0704.1049}}].

\end{thebibliography}\endgroup
\end{document}